\documentstyle[twoside,psfig]{article}

\catcode`\@=11
\long\def\@makefntext#1{
\protect\noindent \hbox to 3.2pt {\hskip-.9pt  
$^{{\eightrm\@thefnmark}}$\hfil}#1\hfill}		

\def\@makefnmark{\hbox to 0pt{$^{\@thefnmark}$\hss}}	
	
\def\ps@myheadings{\let\@mkboth\@gobbletwo
\def\@oddhead{\hbox{}
\rightmark\hfil\eightrm\thepage}   
\def\@oddfoot{}\def\@evenhead{\eightrm\thepage\hfil
\leftmark\hbox{}}\def\@evenfoot{}
\def\sectionmark##1{}\def\subsectionmark##1{}}

\oddsidemargin=\evensidemargin
\newcounter{sectionc}\newcounter{subsectionc}\newcounter{subsubsectionc}
\renewcommand{\section}[1] {\vspace{12pt}\addtocounter{sectionc}{1} 
\setcounter{subsectionc}{0}\setcounter{subsubsectionc}{0}\noindent 
	{\tenbf\thesectionc. #1}\par\vspace{5pt}}
\renewcommand{\subsection}[1] {\vspace{12pt}\addtocounter{subsectionc}{1} 
	\setcounter{subsubsectionc}{0}\noindent 
	{\bf\thesectionc.\thesubsectionc. {\kern1pt \bfit #1}}\par\vspace{5pt}}
\renewcommand{\subsubsection}[1] {\vspace{12pt}\addtocounter{subsubsectionc}{1}
	\noindent{\tenrm\thesectionc.\thesubsectionc.\thesubsubsectionc.
	{\kern1pt \tenit #1}}\par\vspace{5pt}}
\newcommand{\nonumsection}[1] {\vspace{12pt}\noindent{\tenbf #1}
	\par\vspace{5pt}}

\newcounter{appendixc}
\newcounter{subappendixc}[appendixc]
\newcounter{subsubappendixc}[subappendixc]
\renewcommand{\thesubappendixc}{\Alph{appendixc}.\arabic{subappendixc}}
\renewcommand{\thesubsubappendixc}
	{\Alph{appendixc}.\arabic{subappendixc}.\arabic{subsubappendixc}}

\renewcommand{\appendix}[1] {\vspace{12pt}
        \refstepcounter{appendixc}
        \setcounter{figure}{0}
        \setcounter{table}{0}
        \setcounter{lemma}{0}
        \setcounter{theorem}{0}
        \setcounter{corollary}{0}
        \setcounter{definition}{0}
        \setcounter{equation}{0}
        \renewcommand{\thefigure}{\Alph{appendixc}.\arabic{figure}}
        \renewcommand{\thetable}{\Alph{appendixc}.\arabic{table}}
        \renewcommand{\theappendixc}{\Alph{appendixc}}
        \renewcommand{\thelemma}{\Alph{appendixc}.\arabic{lemma}}
        \renewcommand{\thetheorem}{\Alph{appendixc}.\arabic{theorem}}
        \renewcommand{\thedefinition}{\Alph{appendixc}.\arabic{definition}}
        \renewcommand{\thecorollary}{\Alph{appendixc}.\arabic{corollary}}
        \renewcommand{\theequation}{\Alph{appendixc}.\arabic{equation}}
        \noindent{\tenbf Appendix \theappendixc #1}\par\vspace{5pt}}
\newcommand{\subappendix}[1] {\vspace{12pt}
        \refstepcounter{subappendixc}
        \noindent{\bf Appendix \thesubappendixc. {\kern1pt \bfit #1}}
	\par\vspace{5pt}}
\newcommand{\subsubappendix}[1] {\vspace{12pt}
        \refstepcounter{subsubappendixc}
        \noindent{\rm Appendix \thesubsubappendixc. {\kern1pt \tenit #1}}
	\par\vspace{5pt}}

\newcommand{\textlineskip}{\baselineskip=13pt}
\newcommand{\smalllineskip}{\baselineskip=10pt}

\def\eightcirc{
\begin{picture}(0,0)
\put(4.4,1.8){\circle{6.5}}
\end{picture}}
\def\eightcopyright{\eightcirc\kern2.7pt\hbox{\eightrm c}} 

\newcommand{\copyrightheading}[1]
	{\vspace*{-2.5cm}\smalllineskip{\flushleft
	{\footnotesize Modern Physics Letters A #1}\\
	{\footnotesize $\eightcopyright$\, World Scientific Publishing
	 Company}\\
	 }}

\def\abstracts#1#2#3{{
	\centering{\begin{minipage}{4.5in}\footnotesize\baselineskip=10pt
	\parindent=0pt #1\par 
	\parindent=15pt #2\par
	\parindent=15pt #3
	\end{minipage}}\par}} 

\newcommand{\bibit}{\nineit}
\newcommand{\bibbf}{\ninebf}
\renewenvironment{thebibliography}[1]
	{\frenchspacing
	 \ninerm\baselineskip=11pt
	 \begin{list}{\arabic{enumi}.}
        {\usecounter{enumi}\setlength{\parsep}{0pt}     
	 \setlength{\leftmargin 12.7pt}{\rightmargin 0pt} 
         \setlength{\itemsep}{0pt} \settowidth
	{\labelwidth}{#1.}\sloppy}}{\end{list}}

\newcounter{itemlistc}
\newcounter{romanlistc}
\newcounter{alphlistc}
\newcounter{arabiclistc}

\newcommand{\fcaption}[1]{
        \refstepcounter{figure}
        \setbox\@tempboxa = \hbox{\footnotesize Fig.~\thefigure. #1}
        \ifdim \wd\@tempboxa > 5in
           {\begin{center}
        \parbox{5in}{\footnotesize\smalllineskip Fig.~\thefigure. #1}
            \end{center}}
        \else
             {\begin{center}
             {\footnotesize Fig.~\thefigure. #1}
              \end{center}}
        \fi}

\newcommand{\tcaption}[1]{
        \refstepcounter{table}
        \setbox\@tempboxa = \hbox{\footnotesize Table~\thetable. #1}
        \ifdim \wd\@tempboxa > 5in
           {\begin{center}
        \parbox{5in}{\footnotesize\smalllineskip Table~\thetable. #1}
            \end{center}}
        \else
             {\begin{center}
             {\footnotesize Table~\thetable. #1}
              \end{center}}
        \fi}

\def\@citex[#1]#2{\if@filesw\immediate\write\@auxout
	{\string\citation{#2}}\fi
\def\@citea{}\@cite{\@for\@citeb:=#2\do
	{\@citea\def\@citea{,}\@ifundefined
	{b@\@citeb}{{\bf ?}\@warning
	{Citation `\@citeb' on page \thepage \space undefined}}
	{\csname b@\@citeb\endcsname}}}{#1}}

\newif\if@cghi
\def\cite{\@cghitrue\@ifnextchar [{\@tempswatrue
	\@citex}{\@tempswafalse\@citex[]}}
\def\citelow{\@cghifalse\@ifnextchar [{\@tempswatrue
	\@citex}{\@tempswafalse\@citex[]}}
\def\@cite#1#2{{$\null^{#1}$\if@tempswa\typeout
	{IJCGA warning: optional citation argument 
	ignored: `#2'} \fi}}

\def\pmb#1{\setbox0=\hbox{#1}
	\kern-.025em\copy0\kern-\wd0
	\kern.05em\copy0\kern-\wd0
	\kern-.025em\raise.0433em\box0}

\def\fnt#1#2{\footnotetext{\kern-.3em
	{$^{\mbox{\scriptsize #1}}$}{#2}}}

\def\ps@myheadings{%
    \let\@oddfoot\@empty\let\@evenfoot\@empty
    \def\@evenhead{\slshape\leftmark\hfil}
    \def\@oddhead{\hfil{\slshape\rightmark}}
    \let\@mkboth\@gobbletwo
    \let\sectionmark\@gobble
    \let\subsectionmark\@gobble
    }
\font\tenrm=cmr10
\font\tenit=cmti10 
\font\tenbf=cmbx10
\font\bfit=cmbxti10 at 10pt
\font\ninerm=cmr9
\font\nineit=cmti9
\font\ninebf=cmbx9
\font\eightrm=cmr8

\def\qed{\hbox{${\vcenter{\vbox{			
   \hrule height 0.4pt\hbox{\vrule width 0.4pt height 6pt
   \kern5pt\vrule width 0.4pt}\hrule height 0.4pt}}}$}}

\begin{document}
\setlength{\textheight}{7.7truein}  

\thispagestyle{empty}

\normalsize\textlineskip

\setcounter{page}{1}

\copyrightheading{}	

\vspace*{0.88truein}

\centerline{\bf WHICH CONSTITUENT QUARK MODEL IS BETTER?}

\vspace*{0.4truein}
\centerline{\footnotesize FAN WANG}
\baselineskip=10pt
\centerline{\footnotesize\it Department of Physics, Nanjing University, Nanjing, 210093, China}
\vspace*{10pt}

\centerline{\footnotesize Jia-Lun Ping}
\baselineskip=10pt
\centerline{\footnotesize\it Department of Physics, Nanjing Normal University, Nanjing, 210024, China}
\vspace*{10pt}

\centerline{\footnotesize Hou-Rong Pang}
\baselineskip=10pt
\centerline{\footnotesize\it Department of Physics, Nanjing University, Nanjing, 210093, China}
\vspace*{10pt}

\centerline{\footnotesize T.Goldman}
\baselineskip=10pt
\centerline{\footnotesize\it Theoretical Division, Los Alamos National Lab., Los 
Alamos, NM87545, US}

\vspace*{0.228truein}


\vspace*{0.23truein}
\abstracts{A comparative study has been done by calculating the
effective baryon-baryon interactions of the 64 lowest channels
consisting of octet and decuplet baryons with three constituent quark
models: the extended quark gluon exchange model, the Goldstone boson
exchange model and the quark gluon meson exchange hybrid model. We
find that these three models give similar results for 44 channels. 
Further tests of these models are discussed.}{}{}

\vspace*{2pt}

\baselineskip=13pt	        
\normalsize              	
\section{A Debate on Which Constituent Quark Model is Better}
\vspace*{-0.5pt}
\noindent
For low energy quantum chromodynamics(QCD), especially for complicated
quark gluon systems, nonperturbative QCD calculation is still very
difficult, if not impossible, and one must rely on QCD models.  The
constituent quark model is quite successful in understanding hadron
spectroscopy and even hadron interactions. However there has been a
debate on {\it which constituent quark model is better}.$^{1,2,3,4}$
There is a consensus that the constituent quark is a useful effective
degree of freedom for low energy hadron physics. Different authors have
radically different view points on what other proper effective degrees
of freedom may be.

Glozman and Riska proposed that the Goldstone boson is the only other
proper effective degree of freedom (GR).$^1$ Isgur insisted that the
gluon is the proper one (GI).$^2$ Manohar and Georgi argued that in
between the spontaneously broken chiral symmetry and confinement energy
scales, both Goldstone boson and gluon effective degrees of freedom
survive (MG).$^3$ Georgi also did an effective matrix element method
analysis of P-wave baryon spectroscopy with both the quark-Goldstone
boson coupling and the quark-gluon coupling models, finding that the
former fits the data with a smaller $\chi^{2}$, but noted:  ``{\it our
results should not be interpreted, by themselves, as evidence in favor
of the chiral quark model picture over the nonrelativistic quark
model}''.$^3$ K.F. Liu produced a valence lattice QCD result which
supports the Goldstone boson exchange picture, but Isgur pointed out
that this is unjustified.$^{2,4}$ M. Furuichi and K. Shimizu calculated
the baryon spectroscopy further with both the GR and MG quark gluon
meson exchange hybrid models and concluded that "{\it for the moment we
do not have a definite answer which model is better than the other.}"
H.  Garcilazo et al. checked the MG hybrid and GR models too and
concluded that "{\it the similtaneous description of the one and
two-baryon systems still remains an open problem.}"$^5$

If these three constituent quark models (GR, GI, MG) are really quite
different, differences should also appear in their predictions for
baryon-baryon (BB) interactions which are more sensitive to these model
details. Therefore, we did a comparative study of the predictions of
these three models for BB interactions in the lowest 64 channels
consisting of octet and decuplet baryons. Our results show that 44
channels have similar effective BB interactions.

\vspace*{2pt}
\section{A Comparative Study of Three Constituent Quark Models}
\noindent
There are different versions of each model. We choose Glozman's
parametrization of the Goldstone boson exchange, given in Ref.6, as a
typical example of the GR model. The hybrid MG model is widely used in
studies of the BB interaction. We choose the Fujiwara version$^7$ as
the MG example. It is well known that the quark-gluon coupling model,
GI, provides only a repulsive core for the NN interaction but no
intermediate range attraction. This attraction is attributed to
$\sigma$ exchange or to two pion exchange in the hybrid model. Nor does
the GI model include any one pion exchange long range interaction.

We have extended the GI model: We keep as the three quark Hamiltonian
that used by Isgur, i.e., a harmonic confinement plus a color magnetic
hyperfine interaction, making two modifications to extend it to the BB
interaction. First, instead of the usual two centered single quark
orbital wave function, we introduce a delocalized quark orbital wave
function, borrowed from the description of molecular orbitals.  It
incorporates the mutual distortion of the interacting baryons or the
internal excitation of the baryon in the course of interaction, which
enlarges the variational Hilbert space simply. Second, we use a new
parametrization of the confinement potential to take into account
nonperturbative QCD effects which could not have been parametrized by a
two body confinement and color magnetic hyperfine interaction.  It may
not even be possible to check, by hadron spectroscopy, QCD features
such as the three gluon interaction, three body instanton interaction,
etc.  (A detailed discussion can be found in Ref.8.) With these
additions, the extended GI model produces a quantitatively correct NN
intermediate range attraction without invoking $\sigma$ or two pion
exchange. More importantly, this is the unique model which explains the
long known resemblance of the nuclear force to the molecular force
(except for the obvious length and energy scale differences). We use
this extended quark gluon coupling model, which has been called the
quark delocalization and color screening model (QDCSM), as the example
of GI.

We are interested in general features of the BB interactions which
characterize each individual constituent quark model. Therefore, we
calculated the adiabatic effective BB interactions for the 64 lowest
channels consisting of octet and decuplet baryons with these three
constituent quark models.$^9$ Typical results are shown in Figs.1-3.
In Fig.1, two (SIJ=022,-403) of the 17 pure repulsive core channels are
shown. In Fig.2, two weak attraction channels (SIJ=001,010) are
shown. Altogether there are 13 channels where all three models give
such similar effective BB interactions. In Fig.3, two other weak
attraction channels (SIJ=-1[1/2]0,-223) are shown. Altogether there are
14 channels where two of the three models give very similar results;
the third gives a little different, but still similar, effective BB
interaction.

\begin{figure}[htbp] 
\vspace*{13pt}
\centerline{\psfig{file=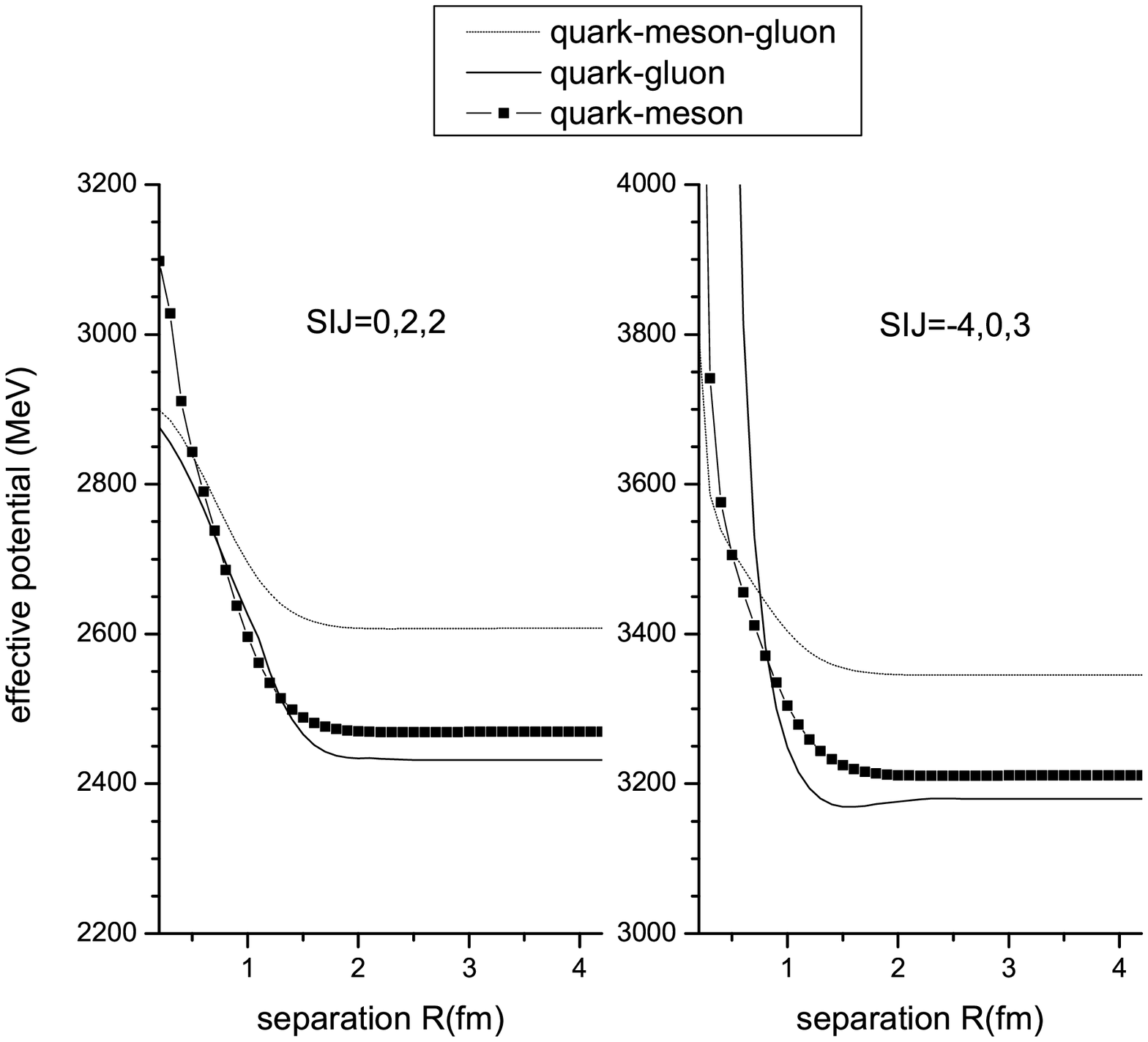,width=6cm,height=4cm}} 
\vspace*{13pt}
\fcaption{Two of the 17 pure repulsive channels.}
\end{figure}

\begin{figure}[htbp] 
\vspace*{13pt}
\centerline{\psfig{file=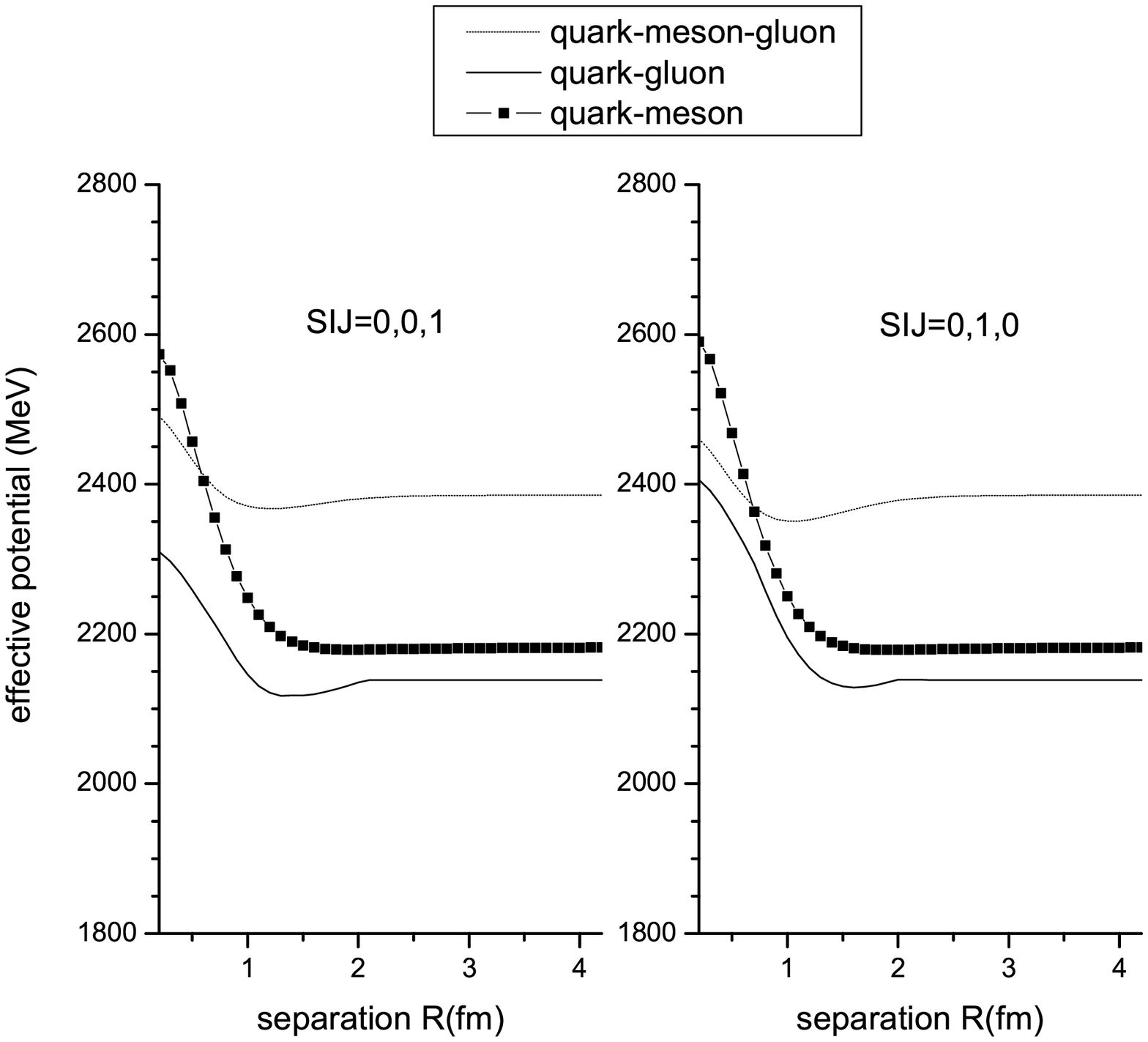,width=6cm,height=4cm}} 
\vspace*{13pt}
\fcaption{Two of the 13 weak attraction channels.}
\end{figure}

\begin{figure}[htbp] 
\vspace*{13pt}
\centerline{\psfig{file=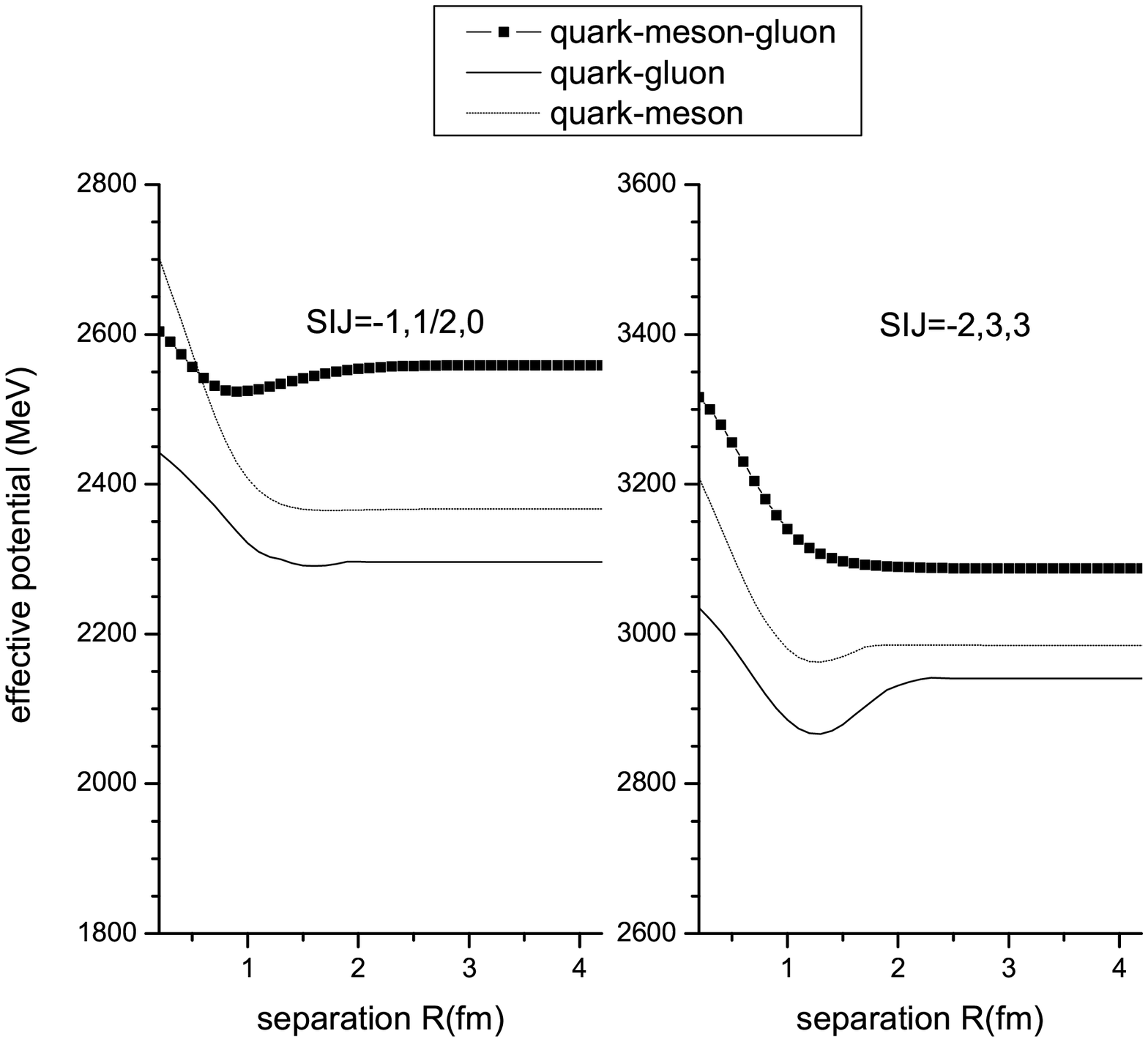,width=6cm,height=4cm}} 
\vspace*{13pt}
\fcaption{Two of the 14 channel where two of the three models are similar 
and one is slightly different. }
\end{figure}

For the H particle channel, the three models all give somewhat
different results but are all weakly attractive with (GR and MG) or
without (QDCSM) a repulsive core. However the dynamical calculations of
the QDCSM and hybrid models yield a state with binding energy around
zero. This is radically different from the naive bag model
estimate$^{10}$ and some earlier hybrid model estimates$^{11}$.

Only in a few channels do the three models give really different
results. For example, the QDCSM predicts a strong attraction in the
di-$\Delta$ (SIJ=003) channel, while the hybrid model predicts a strong
attraction in the di-$\Omega$ (SIJ=-600) channel.

The general features can be summarized as: {\it The three constituent
quark models, although they appear to be very different, give similar
effective BB interactions.}

Returning to hadron spectroscopy, based on the debate$^{1,2,3,5}$, it
is fair to say that while different constituent quark models each have
their own advantages and disadvantages, none gives a perfect
description of hadrons. For example, within the pure valence $q^3$
configuration, no model explains the nucleon spin and flavor structure
and the spin-orbit splitting of the strange and nonstrange mesons and
baryons together.$^{12}$ The GI and MG models have the advantage of a
unified description of the meson and baryon spectrum and the one and
two-baryon systems simultaneously but the disadvantage of predicting
the wrong order for the P and D wave excited baryon states.  The GR
model obtains the correct order for the P and D wave excited states but
loses the unified manner of describing the meson and baryon spectrum
and might find it difficult to describe the one and two baryon systems
simultaneously.

\section{A Tentative Explanation of the Origin of the Similarity}
\noindent
The same pure repulsive core obtained in 17 channels from each of these
three constituent quark models can be understood in that they are due
to Pauli forbidden combinations common to all quark models. The similar
weak attraction obtained in the deuteron channel (SIJ=001) can be
attributed to the adjustment of model parameters because a realistic BB
interaction model should fit the deuteron properties first. However,
there should be physical reasons for the fact that more than 2/3 of the
channels have similar effective BB interactions in all models.

\subsection{Antisymmetrization effect}
\noindent
Antisymmetrization might be one of the physical reasons for the
similarity between the models. To examine this, we neglect the details
of the orbital form and concentrate on the spin-flavor-color structure
of the gluon and Goldstone boson exchange q-q interactions.  (The
hybrid model includes both parts so it does not need to be studied
separately.) The structures are
\begin{equation}
V_{ij}(GE)=-V(GE){\bf \lambda^c_i\cdot\lambda^c_j\sigma_i\cdot\sigma_j},~~~~~~~~
V_{ij}(BE)=-V(BE){\bf \lambda^f_i\cdot\lambda^f_j\sigma_i\cdot\sigma_j}, \label{cf}
\end{equation}
respectively. Recalling the Dirac identity and the antisymmetrization 
condition, we have,
\begin{equation}
{\bf \lambda_i\cdot\lambda_j}=2P_{ij}-2/3,~~~~~~
{\bf \sigma_i\cdot\sigma_j}=2P^s_{ij}-1, \label{p}
\end{equation}
~~~~~~~~~~~~~~~~~~~~~~~~~orbital~~symmetric~~~~~~~~~~~~~~~~~~~~ orbital~~antisymmetric
\begin{equation}
V_{ij}(GE)/V(GE)=4P^f_{ij}+4/3P^s_{ij}+2P^c_{ij}-2/3,~~~~-4P^f_{ij}+4/3P^s_{ij}+2P^c_{ij}-2/3,
\end{equation}
\begin{equation}
V_{ij}(BE)/V(BE)=4P^c_{ij}+4/3P^s_{ij}+2P^f_{ij}-2/3,~~~~-4P^c_{ij}+4/3P^s_{ij}+2P^f_{ij}-2/3.
\end{equation}

Eqs.(3,4) show that due to antisymmetrization, the color dependent
gluon exchange interaction turns out to be flavor dependent and the
flavor dependent Goldstone boson exchange interaction turns to be color
dependent, also. No significant difference appears in the symmetric
orbital case. However, there is a real difference in the antisymmetric
orbital case, where the color and flavor dependent terms have opposite
signs. Therefore we examine this case one step further.

We reexpress these two interactions using Casimir operators,
\begin{equation}
-\sum_{i<j}{\bf \lambda_i\cdot\lambda_j\sigma_i\cdot\sigma_j}=
-4C^{SU(6)}_2+2C^{SU(3)}_2+4/3C^{SU(2)}_2+8N,
\end{equation}
where $C^{SU(n)}_2$ is the second rank Casimir operator of the SU(n)
group and N is the particle number of the system. Next, we obtain the
eigenvalue, $\Delta{E}$, of the gluon and boson exchange interactions
as shown in Table 1, which is calculated for orbital symmetry [6]. The
color part must always be [222] due to color confinement. The
spin-flavor symmetry is restricted to be [33] due to
antisymmetrization. It is then easy to see from Table 1 that the
eigenvalues of these two interactions both decrease with the flavor
symmetry but with slightly different slopes and that they end with the
same minimum at the flavor singlet, namely the H particle state.

\begin{table}[htbp]
\tcaption{$\Delta{E}$ of the gluon and boson exchange interactions.}
\centerline{\footnotesize\smalllineskip
\begin{tabular}{c c c c c c c c}\\
\hline
{[f]} &{[6]} &{[51]} &{[42]} &{[33]} &{[411]}& {[321]} & {[222]}\\
\hline
{[$\sigma$]} &{[33]} &{[42]} &{[51],[33]} & {[6],[42]} & {[42]} & {[51],[42]} &{[33]} \\ \hline
{$\Delta E_{c-s}$} & 48 &80/3 &16,8 &16,8/3 & 8/3 &-4,-28/3 & -24 \\ \hline
{$\Delta E_{f-s}$} & 12 & 8/3 & 0,-8 & 4, -28/3 & -28/3 & -10,-46/3 & -24 \\ \hline
\hline\\
\end{tabular}}
\end{table}

\vspace*{-0.05in}
\subsection{Nonperturbative QCD basis}
\noindent

Having an understanding of the QCD basis of these constituent quark
models is helpful in judging these models. So far, there have been only
approximate nonperturbative QCD "derivations".

Cahill and Roberts "derived" a Goldstone boson-constituent quark
coupling model under the approximation of only keeping the
current-current interaction.$^{13}$ The current quark is dressed to
become the constituent quark in the spontaneously broken chiral QCD
vacuum due to $q\bar{q}$ condensation and the Goldstone boson is
identified as due to quantum fluctuations of the boson field
originating from the $\bar{q}(x)\gamma_5q(y)$ bilocal operator.

Based on this, and the results of Negele et al. on the dilute instanton
model of Shuryak$^{14}$, we developed an approximate QCD "derivation"
of the MG model. The gluon field is separated into two components:  an
instanton part and a quantum fluctuation part. The instanton part is
assumed to be the classical part of the gluon field which induces
spontaneous chiral symmetry breaking and dresses the current quark into
constituent quark, while the quantum fluctuation part of the gluon
field is kept as a perturbative gluon contribution.

In a Schwinger-Dyson equation approach$^{15}$, under a reasonable
truncation, one obtains a dressed quark, gluon propagators and the
dressed quark-gluon vertex which can be translated into an effective
quark-gluon coupling model.

In summary, all three constituent quark models have an arguable basis
in QCD and it is hard to say which one is better inspired by QCD.

\section{An Outlook}
\noindent
Hadron spectroscopy will be improved by new experimental efforts.
Hadronic transition matrix elements are more sensitive to model
details; the internal structure of hadrons tests models more
stringently.  There are already candidates for exotic quark-gluon
systems and pursuit of dibaryons by various groups continues.  New
facilities will provide more precise hyperon-nucleon, and even
hyperon-hyperon, interaction data.  Since each model has its strengths
and weaknesses, the different model approaches to a unified description
of hadron internal structure and hadron interactions are needed to
teach us about hadron physics from different perspectives.

Nonperturbative QCD approaches have shown that the different quark
models may well be related to QCD in different ways. It is highly
desirable to continue this effort to clarify the relation between QCD
theory and QCD models and, in turn, to understand the relation between
different models, especially if they are contradictory, as has led to
strong debate in recent years.  It might be worthwhile to recall that a
similar history occurred in the study of nuclear structure.

\nonumsection{Acknowledgments}
\noindent
This research is supported in part by the National Research Council of
China and in part by the US Department of Energy under contract
W-7405-ENG-36.

\end{document}